\input{aipcheck}

\documentclass[
    ,final            
  ]
  {aipproc}
\usepackage{epstopdf}
\layoutstyle{6x9}
\begin{document}

\title{Future directions for probing two and three nucleon short-range correlations at high energies 
}


\author{Leonid Frankfurt}{
  address={Tel Aviv University, Tel Aviv, Israel}
}
\author{Misak Sargsian}{
  address={Florida International University, Miami, Florida, USA}
}

\author{Mark Strikman}{
  address={Pennsylvania State University, University Park, PA, USA}
}
\begin{abstract}
We summarize recent progress in the studies of the short-rang correlations (SRC) in nuclei in high energy electron and hadron nucleus scattering and suggest directions for the future high energy studies aimed at establishing detailed structure of two-nucleon SRCs, revealing  structure of three nucleon SRC correlations and discovering non-nucleonic degrees of freedom in nuclei.
\end{abstract}

\maketitle

\section{Introduction}
The short-range nucleon correlations (SRC)  for decades were  considered to be important though elusive feature of nuclear structure. These correlations lead to presence of the high momentum tail in the nucleus wave function and they are responsible (in medium and heavy nuclei) for $\ge 60\%$ of the kinetic energy 
of nucleons in the nuclei.

However the SRC are averaged over in  the low energy processes and could be hidden in the parameters of the effective 2N, 3N interaction, effective mass of the nucleon, etc. Such averaging leads to successful description of many low energy phenomena - effective field theories. However it   results in  the wave functions with   very  small high momentum tails and   most of the kinetic energy of nucleons originating from the mean field. 
This is qualitatively different from the realistic nuclear wave functions with SRCs.

Many attempts to probe SRC in the  50's and 60's failed to separate SRC from effects of meson exchange currents, production of $\Delta $ -isobars in the intermediate state, etc. Moreover a no-go theorem was suggested by Amado\cite{Amado} which states that it is in principle impossible to observe the high momentum component of the nucleus.  On the contrary, starting  with our first analysis in 1975 \cite{FS75}  we argued that inconclusive results of the previous searches were  due to insufficient energy/momentum transfer in the studied reactions, leading to complicated structure of primary interaction, enhancement of the final state contributions. We suggested \cite{FS75,FS81} that a way out is to use the processes with large energy-momentum transfer:
\begin{equation}
q_0\ge   1 \mbox{GeV} \gg |V_{NN}^{SR}|,  \vec{q}\ge 1 \mbox{ GeV/c}\gg 2\ k_F.
\end{equation}
Adjusting resolution scale  as a function of the probed nucleon momentum (which is similar to the strategy used in the high energy QCD to probe the parton structure of hadrons) allows to avoid the  no-go theorem \cite{Amado}.  

Consequently, one can use high energy probes to address fundamental questions of microscopic quark-gluon structure 
of nuclei and nuclear forces: 
(i) microscopic origin of intermediate and short-range nuclear forces, (ii) precision of approximation of the bound nucleon wave functions by free nucleon wave functions, (iii) probability and structure of the short-range correlations in nuclei,
(iv) presence and structure of non-nucleonic degrees of freedom in nuclei.

\section{What was learned about short-range correlations in the last few years}
During the last three  years a qualitative progress in the study of  SRC was reached based on the analysis of the high momentum transfer $(e,e')$ Jlab data \cite{Kim1,Kim2}, the (p,2pn) BNL data \cite{eip2,eip3},  the  $(e,e'pp)$ \cite{eip4} and  $(e,e'pn)$ Jlab data \cite{eip4,eip5}.  SRC are not anymore an elusive property of nuclei!

The results of the theoretical analysis of these experimental findings (practically all of  which were predicted before the data were obtained) can be summarized as follows (for a detailed review see \cite{FSS08}):

\begin{itemize}
\item More than $\sim 90\%$ of  all nucleons with momenta $k\ge  \mbox{300 MeV/c}$  belong to two nucleon SRC correlations.

\item Probability for a given proton  with momenta $600\ge  k \ge \mbox{300 MeV/c}$ to belong to a $pn$ correlation is  $\sim$ 18 times larger than to belong to a $ pp$ correlation.

\item Probability for a nucleon to have momentum $\ge$ 300 MeV/c in medium nuclei is  $\sim 25\% $.

\item 2N SRC mostly build of two nucleons not of six quark configurations, or $ \Delta N, \Delta \Delta$ which constitute no more than  $ 20\%$  of the 2N SRCs.
\item Three nucleon SRC are present in nuclei with a significant probability.
\end{itemize}

The findings confirm our predictions based on the study of the structure of SRC in nuclei \cite{FS77,FS81,FS88,FSDS}, and add new information about isotopic structure of SRC. In particular, they confirms our interpretation of the phenomenon of the fast backward hadron emission observed in the 70's-80's in  a number of  high energy $\gamma, \pi,\nu_{\mu},$ p, nucleus - nucleus experiments as to due to SRC. Thus we can now  use information from these experiments for planning new experiments which would allow unambiguous interpretation of $j\ge 3$ SRCs. 

The progress in the studies of SRC which 
led
to the findings summarized above is primarily due to the application  of two concepts:

(a) {\it  Validity of the  partial closure approximation for the inclusive 
(e,e') processes at } $x > 1, Q^2 \ge 1.5 \mbox{GeV}^2$. We   find that in this kinematics only final state interactions (fsi)  at the longitudinal distances $\le $1.2 fm  do not cancel out \cite{FSDS} in the inclusive cross section. At the same time the local fsi's  occur within the SRCs and hence are universal and cancel in the ratios \cite{FS81,FSDS,FSS08}.
 Recently \cite{FSS08} we developed a systematic method to calculate the fsi at large $Q$ using generalized eikonal approximation (GEA)\cite{FSS97}. We find that  interactions of knocked out nucleon with slow nucleons  changes cross section by less than few \%.
 
 (b) {\it Use of hard exclusive processes where a nucleon of SRC is removed instantaneously.}  Such processes probe another quantity sensitive to SRC - nuclear decay function \cite{FS77,FS81,FS88} - probability to emit a nucleon with momentum $k_2$  after removal of a fast nucleon with momentum
  $k_1$, leading to a state with excitation energy $E_r$. In the nonrelativistic limit the decay function can be defined as       
 \begin{equation}
 D_A(k_2,k_1,E_r)=\left|\left<\phi_{A-1}(k_2,...)\left|\delta(H_{A-1}-E_r)a(k_1)\right|\psi_A\right>\right|^2.
 \label{decay}
 \end{equation}
 The general principle which governs the properties of the decay function for large momenta of the removed nucleon is that to release a nucleon "1"  from say two nucleon SRC it is necessary to remove a nucleon "2"  from the same correlation - to perform a work against potential $V_{12}(r)$.
 
 This property of local singular interactions  leads to an operational definition of the SRC: nucleon belongs to SRC if its instantaneous  removal from the nucleus leads to emission of  one or two nucleons which balance its momentum. Following this definition we include in SRC not   only correlations due to the repulsive core but also the ones due to the tensor force interactions. For 2N SRC  one can model decay function as decay of the $NN$ pair moving in mean field (like for spectral function  in \cite{FSCS})    \cite{eip3}. 
 The studies of the spectral and decay functions of $^3$He \cite{eheppn1} reveal 
 both 2N and 3N  SRC and confirm the pattern of the decay of the correlations described above. It is  worth noting 
 here that in the applications to the calculations of the observables it is important to use the light-cone decay functions as they automatically take into account the recoil effects - conservation of the light cone fractions.
 If nucleon momenta are in the nonrelativistic domain the light-cone decay function is close to nonrelativistic decay function.
  
  So far  no methods were developed which would allow  to calculate decay functions for $A >4$. However  the decay function and another interesting characteristics of the nuclear structure,  the  two nucleon momentum distribution in the nuclei,  which can be calculated for a large range of nuclei 
  \cite{Schiavilla,  Alvioli07} are  close   for $\vec{k}_1+\vec{k}_2=0$, $k_1\gg k_F$  though this similarity should break down  with increase of     $ \left|\vec{k}_1+\vec{k}_2\right|$.

  \section{Directions  for study of two nucleon short-range correlations in nuclei}
  
Further 2N correlation studies  in inclusive (e,e') reactions and via study of the decay processes are necessary. In the case of the (e,e') reactions studies of the isospin dependence of the cross sections would complement the studies of the p/n ratio in the decay reactions. 
In the decay processes one needs to focus on the high energy  kinematics with minimal 
fsi
between nucleons of the 2N SRC.
One needs to perform factorization tests for 2N SRC  - namely study the  removal of a nucleon at different Q and by different probes and  demonstrate that decay function is universal. Experience of BNL and Jlab experiments appears to demonstrate that it is easier to move to larger momentum transfer kinematics with hadronic projectiles. Minimal program would be to study  forward - backward correlations for a range of light nuclei $^3$He \&$^4$He in $A(e,eÕpp)$ and $A(e,eÕpn)$ at Jlab at $Q^2=2  \div 4 \mbox{GeV}^2$ and at the  proton facilities 
(J-PARC, GSI)  with protons of energies starting at 6 GeV.  It will be necessary to investigate the A-dependence of the pp/pn ratio, its dependence on momentum of the hit nucleon. This would require statistics which is at least a factor of  100  higher than in the current experiments.
  
 \section{Light-cone wave functions of nuclei}
  There is a price to  pay for use of high energy processes:  high energy process develops along the light cone (LC).  Hence similar to the perturbative QCD the amplitudes of the processes are expressed through LC wave functions of the probed system. At the same time  for the nonrelativistic 
 momentum component in nuclei and for 2N SRC correspondence with nonrelativistic wave functions is unambiguous and  rather simple due to the angular condition  \cite{FS76,FS81,FS88}.         
  Many features of nonrelativistic quantum mechanics hold - number of degrees of freedom and dynamic variables remain the same,   though the relations between wave functions and the amplitudes become somewhat different.   At the same time logic of quantum mechanics does not map easily  to the language based on the  
   virtual particles - transformational vacuum pairs lead to  extra degrees of freedom. For example, in the LC model  the deuteron is described by  two (S- and D-) wave functions while in the Bethe-Salpeter model one has to introduce  four (for one nucleon off shell) wave functions.
  
 One can define the single nucleon LC wave function $\rho_A(\alpha,p_t)$ where  $\alpha $ is the LC fraction carried by a nucleon scaled to A ($\sum_{i=1}^A \alpha_i =A$).  One can also define LC spectral function,  $\rho_A^N(\alpha, p_t,M_{rec}^2)$ which 
 after account of the angular momentum conservation (angular condition)
   depends on two variables like the nonrelativistic  spectral function. The LC spectral function enters into  description of     $(e,e')$ at large $Q^2$ and $x> 1$. Early validity of the closure for $x < 2$ leads to possibility to use  the relation $\int d^2M^2_{rec} \rho_A^N(\alpha, p_t,M_{rec}^2) =\rho^N_A(\alpha,p_t)$. Also for production of  fast backward nucleon in the high energy hadron - nucleus  scattering we find 
  \begin{equation}
  {d\sigma^{h+A\to N +X}\over {d\alpha d^2p_t\over \alpha}}=\kappa_h A\sigma_{in}^{hN} \rho^N_A(\alpha,p_t),
  \end{equation}
where  factor    $\kappa_h \sim 1$  which is a weak function of $\alpha$        accounts for local screening effects in the interaction of the projectile with the SRC.

Since the $NN$ interaction is sufficiently singular for large nucleon momenta, $\rho^N_A(\alpha,p_t)$ can  be expanded over contributions of j-nucleon correlations as \cite{FS88}:
\begin{equation}
\rho_A^N(\alpha >1.3,p_t)=\sum_{j=2}^A\rho_j(\alpha,p_t),
\end{equation}
where $\rho_j(\alpha,p_t) (j-\alpha)^{n(j-1)+j-2}$ and $ \rho_j(\alpha,0)\propto (2-\alpha)^n$. Note that the LC density matrix behavior at large $\alpha \ge 2$ is determined by multinucleon correlations while  in the nonrelativistic case the $k\to \infty$ asymptotic of the the momentum distribution, n(k)  is due to two nucleon correlations.

   \section{Directions  for study of three nucleon short-range correlations in nuclei}
   
   Current evidence for the presence of the three nucleon correlations comes from the study of the production of fast backward nucleon production at $\alpha\ge 1.5$ and  from the scaling of the ratios of the  $(e,e')$ 
cross sections at $3> x > 2$ for $^3$He/$^4$He \cite{FS88} and for a wider range of nuclei in \cite{Kim1}.  
Also, the theoretical studies of the three nucleon wave functions have revealed configurations in the spectral and decay functions which are determined by the three nucleon correlations \cite{eheppn1,FSS08}.

Currently a dedicated  Jlab experiment is approved to study 3N correlations for $x> 2$. 
The range of the $Q^2$ will be rather limited and accuracy of the scaling relations in the kinematics of the experiment requires further theoretical studies.

\begin{figure}[ht]
\centering\includegraphics[height=4cm,width=8cm]{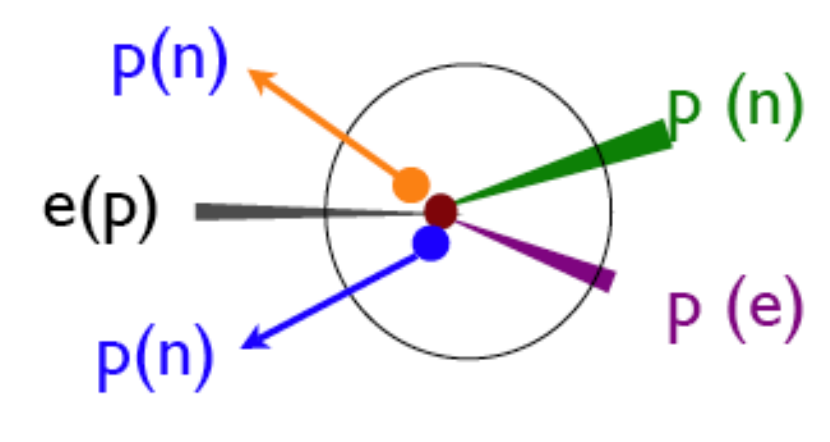}
\caption{Quasielastic scattering of electron (proton) off  3N SRC with production of two backward nucleons.}
\label{3N}
\end{figure}
One of the possible ways to study 3N correlations would be to investigate processes where a projectile knocks out a forward moving nucleon of the nucleus and one detects two nucleons emitted backward (Fig.1).

One expects 
(i)
$\alpha_{1\, Back.Nucl}+ \alpha_{2\, Back.Nucl}\alpha_{3\, Forw.Nucl} \approx 3$, (ii) similar rate of $ppn$ and $nnp$ events (when corrected for a different elementary cross section of projectile proton / neutron interaction) and much smaller rate   for $nnn$ and $ppp$ events due to dominance of the $I=0$ 2N correlations which build up 3N correlations (similar trend is expected due to the contribution of the three nucleon forces), (iii) strong correlation of the transverse momenta of the backward nucleons - the transverse angle $\psi$ between directions of the transverse momenta $k_{1t} $ and $k_{2t}$ of two backward nucleons. Most of the events  should have  $\psi \sim \pi$, while the yield  for  $\psi \sim 0$ should be strongly suppressed because  for the parallel emission of two backward nucleons the  LC  invariant mass for the three nucleon system is much larger.

The last expectation is in a qualitative agreement with the studies of the inclusive process $p+ A\to pp +X$ at $p_{inc}=7.5 \ $GeV/c where correlation function
$$R_2={1\over \sigma^{in}_{pA}}{d\sigma(p+A\to pp+X)/d^3p_1d^3p_2 \over d\sigma(p+A\to p+X)/d^3p_1 d\sigma(p+A\to p+X)/d^3p_2}$$ was measured for a fixed azimuthal angle $\theta = 120^o$. $R_2 $ was found to drop strongly for  the angles $\psi$ deviating from $\pi$.   A detailed analysis of these data will be presented elsewhere.

  \section{How to discover the structure of nonnucleonic baryonic degrees of freedom in nuclei}

What is the 
domain 
in momentum space where  description of NN correlations in terms of nucleonic degrees of freedom maybe justified? If this momentum range is small  and too many states should be included in the decomposition over hadronic states,  the Fock representation 
$$\left|D\right>=\left|NN\right>+\left|NN\pi\right>+\left|\Delta \Delta\right>+ \left|NN\pi\pi\right>+ ...$$
would be useless. 

In view of limited knowledge of details of the dynamics of the off shell  $NN$ interactions we have to use experimental  information on NN interactions at energies below few GeV and the chiral dynamics combined with the following general quantum mechanical principle - {\it relative magnitude of different components in the wave function should be similar to that in the NN scattering at the energy corresponding to off-shellness of the component}.

Important simplification of the dynamics  is due to  the structure of the final states in NN interactions: direct pion production is suppressed for a wide range of energies due to chiral properties of the NN interactions \cite{FS88}:
\begin{equation}
{\sigma(NN\to NN\pi)\over \sigma(NN\to NN)} \approx {k_{\pi}^2 \over 16 \pi^2F_{\pi}^2}, \, F_{\pi}=\mbox{94 MeV.}
\end{equation}
Consequently, the main inelasticity for NN scattering for $T_p \le \mbox{1 GeV}$ is single  $\Delta$-isobar production which is forbidden in the deuteron channel where inelastic threshold corresponds to production of two $\Delta$-isobars 
i.e.
to $k_N=\sqrt{m_{\Delta}^2-m_N^2} \approx 800 \, \mbox{MeV/c} \,$ ! For the I=1 channel the single $\Delta$ production threshold corresponds to $k_N\approx$ 500 MeV/c. (Note that the correspondence argument  for connection between wave functions of bound state  and continuum is not applicable for the cases when the probe interacts with rare configurations in the bound nucleons due to the presence of an additional scale.)

To summarize: pn  and pp correlations are predominantly build of nucleons   
with an
10$\div $ 20 \% accuracy. Thus  exotic components (6q, $\Delta$) -isobars)  should be corrections even in SRC where energy scale is larger than for the mean field configurations and internucleon distances are $<$ 1.2 fm.

The EMC effect for $0.3<x<0.7$  unambiguously indicates presence of non-nucleonic degrees of freedom in nuclei.   (Claims to the  opposite are due to the violation  of  baryon  or energy-momentum conservation or both). Also the Drell-Yan experiments  observe no  enhancement of the pion field. Hence it appears that  looking for exotic baryonic degrees of freedom is the most promising strategy.

According to the correspondence logic for large nucleon momenta admixture of  
configurations 
containing $\Delta$-isobars should not be too small as the energy denominators for $NN$ and $N\Delta$ intermediate states become comparable. The hard probes should resolve intermediate states with $\Delta$'s which are usually hidden in the definition of the $NN$ potential. 

Few possible  strategies for looking for baryonic non-nucleonic degrees of freedom are

(a)  looking for spectator $\alpha_{\Delta} \ge 1$ production. 
Selection of events with $ x> 0.1$ leads to a very strong suppression of two step mechanisms \cite{FS81}. In fact the best limit  on the probability  of $\Delta^{++}\Delta{-}$ component in the deuteron  $<  0.2\%$ comes from the neutrino experiment performed in this kinematics \cite{Allasia}. 
The study  of $\Delta$ production in the $\alpha \ge 1$ kinematics provides the only  possible  experimental evidence for presence of the spectator $\Delta$'s in nuclei. The experiment was performed using  DESY AGRUS  data  on electron - air scattering at $E_e$=5 GeV \cite{Degtyarenko}.
The   $\Delta^{++}/p, \Delta^{0}/p$ ratios  for the same light cone fraction $\alpha$ were measured:\begin{eqnarray} 
{\sigma (e+A \to \Delta^0 + X)\over 
\sigma (e+A \to \Delta^{++} + X)}=0.93\pm 0.2 \pm 0.3, \\   
\nonumber
 \, \, {\sigma (e+A \to \Delta^{++} + X)\over 
\sigma (e+A \to p + X)}=(4.5\pm 0.6 \pm 1.5)\cdot 10^{-2}.
\end{eqnarray}
Clearly, CLAS is in a good position to check this result and get by far superior data.

(b) Knock out of $\Delta^{++} $ isobars in electron scattering at sufficiently large $Q^2$. Examples of good channels are $e+^2\mbox{H}\to e +  \mbox{fast} \,\, \Delta^{++} +  \mbox{slow} \,\,  \Delta^{-}, \, \, e+^3\mbox{He}\to e +  \mbox{fast} \,\, \Delta^{++} +  \mbox{slow} \,\, \mbox{ nn}$.

One can perform similar studies using hadronic projectiles in the kinematics where projectile experiences a large angle elastic scattering off a constituent of the nucleus, for example p+A $\to \Delta^{++}$+p +(A-1). An important tool for analysis of this process and for separation of the one step and two step processes is   measurement of the $\alpha_{\Delta}$ distribution which is much broader for the one step processes of scattering off exotic components off the nucleus.
 
 \section{Conclusions}

Impressive experimental progress of the last three years - discovery of strong short range correlations in nuclei with strong dominance of I=0 SRC,  proves validity of the strategy of using  high momentum transfer processes for probing SRCs. It provides a solid basis for further experimental studies. 

There are many  theoretical challenges  in the studies of the SRC including  calculation of the decay functions, study of the  isotopic effects for SRC, calculating admixture of isobars, study of the  relativistic effects.
Further investigations are also necessary of   the fsi dynamics. This would allow to find optimal kinematics for probing SRCs, and understanding the  role of the color transparency effects. The Generalized Eikonal Approximation (GEA) (see review in \cite{misak} ) provides a   good starting point for such analyses. GEA allows also   include in a consistent way fsi of the produced isobars. It would be important to   perform  experimental tests  of GEA in the kinematics where  isobar fsi's are maximal.

Several experiments to probe SRC are under way/ been planned for 12 GeV. One would  need also  more coherence in the program and  complementary studies using hadron beams. 

This work is supported by DOE grants under contract DE-FG02-01ER-41172 and 
DE-FG02-93ER40771 as well as by the Israel-USA Binational Science Foundation 
Grant.


\begin{thebibliography}{07}
\bibitem{Amado}R.~D.~Amado, Phys.\  Rev. \  {C 19}, 1473 (1979).
\bibitem{FS75}  L.~L.~Frankfurt and M.~I.~Strikman, Proceedings of 10 Winter School  LPNI on Nuclear and Particle 
Physics. Leningrad v2 (1975) 3; Preprint LPNI 173 (1975).

\bibitem{FS81}  L.~L.~Frankfurt and M.~I.~Strikman, Phys.\ Rept.\  {\bf 76} (1981) 215.


\bibitem{Kim1} K.~S.~Egiyan {\it et al.}  [CLAS Collaboration], Phys.\ Rev.\ Lett.\  {\bf 96} (2006) 082501 
\bibitem{Kim2}   K.~S.~Egiyan {\it et al.}  [CLAS Collaboration],Phys.\ Rev.\  C {\bf 68} (2003) 014313 


\bibitem{eip2} A.~Tang {\it et al.},   Phys.\ Rev.\ Lett.\  {\bf 90} (2003) 042301 
\bibitem{eip3}E.~Piasetzky, M.~Sargsian, L.~Frankfurt, M.~Strikman and J.~W.~Watson,
               Phys.\ Rev.\ Lett.\  {\bf 97} (2006) 162504 
\bibitem{eip4}R.~Shneor {\it et al.}  [Jefferson Lab Hall A Collaboration], Phys.\ Rev.\ Lett.\  
              {\bf 99} (2007) 072501 [arXiv:nucl-ex/0703023].
\bibitem{eip5}R.~Subedi {\it et al.}R. Subedi et al., Science 320 (2008) 1476.
\bibitem{FSS08}L.~L.~Frankfurt, M.~M.~Sargsian and M.~I.~Strikman, arXiv:0806.4412 [nucl-th].
\bibitem{FS77}L.~L.~Frankfurt and M.~I.~Strikman,  Phys.\ Lett.\  B {\bf 69} (1977) 93.

\bibitem{FS88}  L.~L.~Frankfurt and M.~I.~Strikman, Phys.\ Rept.\  {\bf 160} (1988)  235.

\bibitem{FSDS} L.~L.~Frankfurt, M.~I.~Strikman, D.~B.~Day and M.~Sargsian, Phys.\ Rev.\  C {\bf 48}  (1993) 2451.
\bibitem{FSS97}L.~L.~Frankfurt, M.~M.~Sargsian and M.~I.~Strikman,
               Phys.\ Rev.\  C {\bf 56}(1997) 1124.

\bibitem{FSCS}C.~Ciofi degli Atti, S.~Simula, L.~L.~Frankfurt and M.~I.~Strikman,       
               Phys.\ Rev.\  C {\bf 44}(1991) 7.
              \bibitem{eheppn1}M.~M.~Sargsian, T.~V.~Abrahamyan, M.~I.~Strikman and L.~L.~Frankfurt,
                 Phys.\ Rev.\  C {\bf 71}, 044614 (2005) [arXiv:nucl-th/0406020].
  \bibitem{Schiavilla}R.~Schiavilla, R.~B.~Wiringa, S.~C.~Pieper and J.~Carlson,  
        Phys.\ Rev.\ Lett.\  {\bf 98}, 132501 (2007) [arXiv:nucl-th/0611037]. 
                
      \bibitem{Alvioli07}M.~Alvioli, C.~Ciofi degli Atti and H.~Morita,
  Phys.\ Rev.\ Lett.\  {\bf 100}, 162503 (2008) [arXiv:nucl-th/0709.3989].
\bibitem{FS76}   L.~L.~Frankfurt and M.~I.~Strikman,
  Phys.\ Lett.\  B {\bf 65}, 51 (1976).
\bibitem{Bayukov} Sov.\ J.\ Nucl.\ Phys. \ {\bf  44} 263 (1986).
\bibitem{Allasia}
  D.~Allasia {\it et al.},
  Phys.\ Lett.\  B {\bf 174}, 450 (1986).

\bibitem{Degtyarenko}
  P.~V.~Degtyarenko, Yu.~V.~Efremenko, V.~B.~Gavrilov and G.~A.~Leksin,
Preprint ITEP-90-12 (1990).

\bibitem{misak}
M. M. Sargsian, Int. \ J. \ Mod. Phys. \ E {\bf 10}, 405 (2001). 


\end{thebibliography}
\end{document}